\documentclass[floats,floatfix,showpacs,amssymb,prd,twocolumn,superscriptaddress,nofootinbib,nolongbibliography,reprint]{revtex4-2}

\usepackage{amssymb,amsmath,verbatim,mathtools,needspace,enumitem,etoolbox,graphicx,physics,microtype,afterpage,xspace,tabularx,lmodern,multirow}
\usepackage{gensymb}
\usepackage[dvipsnames, usenames]{xcolor}
\definecolor{linkcolor}{rgb}{0.0,0.3,0.5}
\usepackage[unicode, colorlinks=true, linkcolor=linkcolor, citecolor=linkcolor, filecolor=linkcolor, urlcolor=linkcolor, linktocpage, breaklinks]{hyperref}
\usepackage[all]{hypcap}
\usepackage[T1]{fontenc}
\usepackage[utf8]{inputenc}
\usepackage[usenames,dvipsnames]{xcolor}
\hypersetup{colorlinks=true,citecolor=romared,linkcolor=romared,urlcolor=romared}

\definecolor{romared}{RGB}{142,0,28}

\newcommand{\be}{\begin{equation}}
\newcommand{\ee}{\end{equation}}

\def\be{\begin{equation}}
\def\ee{\end{equation}}
\newcommand{\beq}{\begin{eqnarray}}
\newcommand{\eeq}{\end{eqnarray}}

\usepackage{aas_macros}
\usepackage{makecell}
\usepackage{soul}
\usepackage{comment}
\usepackage[nolist,nohyperlinks]{acronym}

\allowdisplaybreaks

\begin{document}

\pagenumbering{arabic}

\title{Reply to Comment on ``Analysis of Ringdown Overtones in GW150914''}

\author{Gregorio Carullo}
\affiliation{Niels Bohr International Academy, Niels Bohr Institute, Blegdamsvej 17, 2100 Copenhagen, Denmark}
\author{Roberto Cotesta}
\affiliation{Department of Physics and Astronomy, Johns Hopkins University,
3400 N. Charles Street, Baltimore, Maryland, 21218, USA}
\author{Emanuele Berti}
\affiliation{Department of Physics and Astronomy, Johns Hopkins University,
3400 N. Charles Street, Baltimore, Maryland, 21218, USA}
\author{Vitor Cardoso}
\affiliation{Niels Bohr International Academy, Niels Bohr Institute, Blegdamsvej 17, 2100 Copenhagen, Denmark}
\affiliation{CENTRA, Departamento de F\'{\i}sica, Instituto Superior T\'ecnico -- IST, Universidade de Lisboa -- UL,
Avenida Rovisco Pais 1, 1049-001 Lisboa, Portugal}
\maketitle

\noindent
\textbf{\em Summary} -- The ``Comment''~\cite{Isi:comment} discusses the impact on our analysis (the ``Article'')~\cite{Cotesta:2022pci} of: 
(i) a spurious shift in the discretized time axis within \texttt{pyRing}~\cite{pyRing}; 
(ii) too short of an analysis segment ($T=0.1$\,s instead of $T\geq 0.2$\,s). 

In this Reply, we repeat the analysis including these two effects, alleviating some of the previous discrepancies observed with respect to Ref.~\cite{Isi:2022mhy}.
However, even after accounting for the aforementioned remarks, the Bayesian log-evidence for an overtone remains negative, supporting the main findings of the Article.
This overall conclusion is in agreement with two independent reanalyses, Refs.~\cite{Finch:2022ynt, Crisostomi:2023tle}, and with Fig.~1 of Ref.~\cite{Isi:2022mhy}, by the authors of the Comment. 
The latter shows how at times consistent with the peak ($t \sim 2M$ in the units of that figure), the significance does not reach $2 \sigma$.
The fact that the significance is much lower than the original $3.6 \sigma$ estimate~\cite{Isi:2019aib} in all of these works is a consequence of the uncertainty in $t_{\rm start}$, not included in Ref.~\cite{Isi:2019aib}. 
As we pointed out in the Article (following ideas introduced in Ref.~\cite{Carullo:2019flw}), accounting for this uncertainty is crucial to obtain unbiased results.
This has proven to be important also in the recent analysis of Ref.~\cite{Siegel:2023lxl}.

Different choices of sampling rate (2048 Hz in Ref.~\cite{Isi:2022mhy}, 16384 Hz in the Article), and noise estimation methods (in the Article taken to be the same as in the original analysis of Ref.~\cite{Isi:2019aib}) are behind remaining differences.
As shown in the Comment, the two codes give in fact indistinguishable output, if given identical inputs.
The aforementioned remaining technical differences give rise to appreciably different parameter estimates, highlighting how at current detector sensitivity and when targeting short-lived components, such analyses can be significantly affected by technical details.

\begin{figure*}[thbp]
\centering
\includegraphics[width=\textwidth]{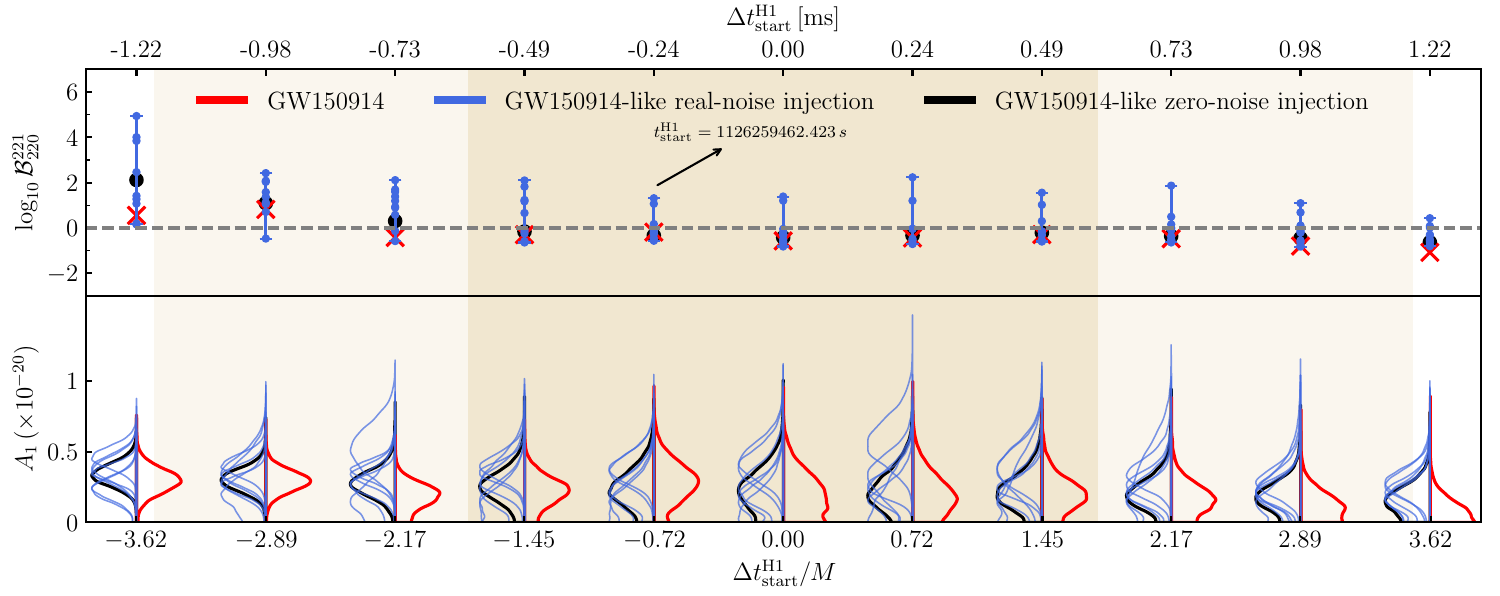}
\caption{
Same as Fig.2 in the Article, but using the updated settings described above.
The results were obtained with the \texttt{pyRing} commit \texttt{6d7d0e8d}.
}
\label{fig:BF_and_amp_N1_vs_N0_vary_time}
\end{figure*}

\noindent
\textbf{\em Results} -- Figure~\ref{fig:BF_and_amp_N1_vs_N0_vary_time} of this Reply updates Fig.~2 of the Article. Taking into account the remarks in the Comment, we have increased the analysis duration from $T=0.1$\,s to $T=0.2$\,s, and removed the time axis shift.
The posteriors show a small shift in favor of an overtone at the $t_{\rm peak}$ value employed by Ref.~\cite{Isi:2019aib}, confirming an improved agreement with the latter analysis, but significant railing is still observed when the $t_{\rm peak}$ uncertainty is taken into account.
The Bayesian log-evidence for an overtone is negative throughout almost the entire time range.
Comparing to Fig.~2 in the Comment, agreement is also observed in the injection set: noise (gray line there) frequently increases the overtone evidence, as expected given the current significance.

Although differing in the technical implementation and in certain analysis inputs, both our results and the ones of Ref.~\cite{Isi:2022mhy} broadly agree with the reanalysis of Ref.~\cite{Finch:2022ynt}.
Here, the authors employ frequency-domain methods and marginalize over the uncertainty in $t_{\rm peak}$, reporting \textit{negative} Bayesian log-evidence for an overtone and a significance of $1.8 \sigma$.
Similar conclusions were found in Ref.~\cite{Crisostomi:2023tle}.
The recent re-analysis of Ref.~\cite{wang2023frequencydomain} gives results more in line with the Comment, compared to ours.
This is expected because they employ a noise estimation method  similar to the one adopted in the reanalysis of the Comment.

Below, we address the specific technical points raised in the Comment.

\noindent
\textbf{\em Time axis shift} -- 
In Refs.~\cite{Isi:comment,Isi:2019aib,Cotesta:2022pci,Isi:2022mhy} the starting time of the analysis, $t_{\rm start}$, is fixed and assumed to agree with $t_{\rm peak}$, the time at which the gravitational-wave strain has a peak.
The internal selection of $t_{\rm start}$ in \texttt{pyRing} conservatively selects the time-stamp immediately following the requested $t_{\rm start}$.
The time axis mislabeling pointed out in the Comment implies a misidentification between the closest data time-stamp and $t_{\rm start}$.
At the sampling rate used in the Article, this approximation leads to a shift of only $\sim 0.06\, \rm ms$, as quantified in~\cite{pyRing_issue}.
The statistical uncertainty in the value of $t_{\rm peak}$ determined from the data is $\sim 2.5\, \rm ms$, implying that the shift (i.e. the ``software bug'' discussed in the Comment) has an overall negligible impact on the analysis, being $\sim 40$ times smaller than this uncertainty.
This is confirmed by the bottom panel of Fig.~1 in the Comment.
The above mislabeling has been removed since \texttt{pyRing v2.1.0}.
Finally, the  $0.24\, \rm ms$ shift in Fig.~2 of the Article, mentioned in the Comment, was applied in post-processing for better comparison with Ref.~\cite{Isi:2019aib} when using a lower sampling rate. This shift did not enter our analysis. 
To facilitate comparison with Ref.~\cite{Isi:2022mhy}, we do not apply any time shift in Fig.~\ref{fig:BF_and_amp_N1_vs_N0_vary_time} of this Reply.

\noindent
\textbf{\em Analysis duration} -- While the signal power of the unfiltered GW150914 ringdown signal after $0.1\, \rm s$ is negligible, this may not be the case for the whitened signal entering the time-domain likelihood, which (depending on both the noise and model features) can be significantly stretched due to the overlap of the mode's spectrum with the detector's spectral lines.
Increasing the analysis duration to $T=0.2\, \rm s$ produces a small increase in the inferred amplitude of the searched overtone, as shown in the bottom panel of Fig.~1 of the Comment, and in Fig.~\ref{fig:BF_and_amp_N1_vs_N0_vary_time} of this Reply.
However, while this increase is appreciable at $t_{\rm start}$ of Fig.~1 in~\cite{Isi:comment} (corresponding to $-0.72M$ on our time axis), Fig.~\ref{fig:BF_and_amp_N1_vs_N0_vary_time} below still shows significant railing against zero when $t_{\rm start}$ is set to nearby values.

\noindent
\textbf{\em Remaining differences} -- 
In Ref.~\cite{Isi:2022mhy}, the auto-correlation function (ACF) was estimated using a Fourier (FFT) method (which is closer to the standard method used in LIGO-Virgo-Kagra analyses within the \texttt{pyRing} pipeline~\cite{Carullo:2019flw, LIGOScientific:2020tif}), while the Article used the same ACF estimation method of the original analysis Ref.~\cite{Isi:2019aib}, obtained through a direct correlation. 
We test the validity of the direct correlation vs FFT estimates by dividing 32\,s of data surrounding the event into stretches 200\,ms long, whitening each of these.
A Kolmogorov-Smirnov test against the gaussian hypothesis reports less than $1\%$ of outliers at $1\%$ significance with both methods for the 32\,s surrounding the event, confirming the statistical validity of both noise estimates within this test.

We also test for putative numerical instabilities in our analysis. 
We evaluate the covariance matrix conditioning number, and compute the time-domain likelihood using supposedly more robust inversion methods (based on the Cholesky decomposition combined with back-substitution or the Levinson recursion), instead of a direct inversion.
We obtain identical likelihoods up to double precision, implying that no numerical instabilities are present.

When repeating the analysis of Fig.~\ref{fig:BF_and_amp_N1_vs_N0_vary_time} using a sampling rate of 2048 Hz\footnote{The higher sampling rate in our analysis was chosen to obtain the maximum available precision on the start time.} and an FFT ACF estimate, we find results in closer agreement with Ref.~\cite{Isi:comment}, confirming that the two analysis softwares (\texttt{ringdown} and \texttt{pyRing}) achieve close to identical outputs when they are given the same inputs.
In this case, the overtone significance is larger, but significant railing is still present for $\Delta t^{H1}_{\rm start} = 0.72, 1.45\, M$, and the Bayesian evidence is always not significant within the $t_{\rm peak}$ uncertainty at $1\sigma$.
This is consistent with the substantial railing evident even in the results of Ref.~\cite{Isi:2022mhy}, as shown in Fig.~4 of the Article. Note that the railing is more apparent when we plot histograms of the samples, instead of the bands shown in Fig.~2 of the Comment.

\noindent
\textbf{\em Additional remarks on the Comment} -- 
The Comment still fails to address the key observation of the Article, namely, that accounting for the statistical uncertainty in $t_{\rm peak}$ (not incorporated in the original analysis of Ref.~\cite{Isi:2019aib}) is necessary to obtain unbiased results.
To detect a mode, the evidence should hold throughout the support of the distribution of $t_{\rm peak}$, not at a single (arbitrary) value.
A positive outcome emerged from the present debate is that new analyses by different groups have instead started to take into account this crucial uncertainty~\cite{Finch:2022ynt,Siegel:2023lxl}.

It should also be noted how, for clarity, the time axis of Fig.~2 of the Comment should be centered around the best estimates of the peak time derived by the Comment's authors in Ref.~\cite{Isi:2022mhy}, i.e. $\sim 2M$ later than the original estimate $t_{\rm ref}$ used in Ref.~\cite{Isi:2019aib}. 
At that point, the significance has dropped substantially: as shown in Fig.~7 of Ref.~\cite{Isi:2022mhy}, even at this best estimate for $t_{\rm peak}$, the log-Bayes factor is below unity.
As we show with the injection of numerical relativity waveforms (for which we know the true value of $t_{\rm peak}$), analyzing the signal at times earlier than $t_{\rm peak}$ can only artificially increase the evidence for an overtone, as the model is trying to fit the merger part of the waveform.
In summary, even using the posteriors computed by the authors of the Comment, the evidence for a second mode is not significant (see also Fig.~4 of the Article). 

\noindent
\textbf{\em Theoretical considerations} -- Beyond considerations of statistical evidence, recent studies of the overtones~\cite{Baibhav:2023clw, Nee:2023osy,2023arXiv231004489H} and of nonlinear mode excitation in numerical simulations~\cite{Cheung:2022rbm,Mitman:2022qdl,Khera:2023lnc} have clearly shown that the test presented in Ref.~\cite{Isi:2019aib} is not relevant for the observational investigation of black hole spectra. 
The claimed ``overtone'' does not correspond to the physical excitation of a true quasinormal mode close to the peak. It is merely a phenomenological term fitting away nonlinearities near the peak of the radiation. 
Note that phenomenological tests in the merger-ringdown regime have already been conducted at the time of the first gravitational-wave detection~\cite{LIGOScientific:2016lio}.
Even if the significance of a subdominant damped exponential were higher, as claimed in previous analyses, it would be wrong to identify this second exponential with the first overtone and to use it for spectroscopy tests.
This argument alone implies that the analysis of Ref.~\cite{Isi:2019aib} should not be interpreted as a test of the ``no-hair theorem.''

\bibliography{References_arXiv}

\begin{thebibliography}{19}%
\makeatletter
\providecommand \@ifxundefined [1]{%
 \@ifx{#1\undefined}
}%
\providecommand \@ifnum [1]{%
 \ifnum #1\expandafter \@firstoftwo
 \else \expandafter \@secondoftwo
 \fi
}%
\providecommand \@ifx [1]{%
 \ifx #1\expandafter \@firstoftwo
 \else \expandafter \@secondoftwo
 \fi
}%
\providecommand \natexlab [1]{#1}%
\providecommand \enquote  [1]{``#1''}%
\providecommand \bibnamefont  [1]{#1}%
\providecommand \bibfnamefont [1]{#1}%
\providecommand \citenamefont [1]{#1}%
\providecommand \href@noop [0]{\@secondoftwo}%
\providecommand \href [0]{\begingroup \@sanitize@url \@href}%
\providecommand \@href[1]{\@@startlink{#1}\@@href}%
\providecommand \@@href[1]{\endgroup#1\@@endlink}%
\providecommand \@sanitize@url [0]{\catcode `\\12\catcode `\$12\catcode
  `\&12\catcode `\#12\catcode `\^12\catcode `\_12\catcode `\%12\relax}%
\providecommand \@@startlink[1]{}%
\providecommand \@@endlink[0]{}%
\providecommand \url  [0]{\begingroup\@sanitize@url \@url }%
\providecommand \@url [1]{\endgroup\@href {#1}{\urlprefix }}%
\providecommand \urlprefix  [0]{URL }%
\providecommand \Eprint [0]{\href }%
\providecommand \doibase [0]{https://doi.org/}%
\providecommand \selectlanguage [0]{\@gobble}%
\providecommand \bibinfo  [0]{\@secondoftwo}%
\providecommand \bibfield  [0]{\@secondoftwo}%
\providecommand \translation [1]{[#1]}%
\providecommand \BibitemOpen [0]{}%
\providecommand \bibitemStop [0]{}%
\providecommand \bibitemNoStop [0]{.\EOS\space}%
\providecommand \EOS [0]{\spacefactor3000\relax}%
\providecommand \BibitemShut  [1]{\csname bibitem#1\endcsname}%
\let\auto@bib@innerbib\@empty
\bibitem [{\citenamefont {Isi}\ and\ \citenamefont {Farr}(2023)}]{Isi:comment}%
  \BibitemOpen
  \bibfield  {author} {\bibinfo {author} {\bibfnamefont {M.}~\bibnamefont
  {Isi}}\ and\ \bibinfo {author} {\bibfnamefont {W.~M.}\ \bibnamefont {Farr}},\
  }\href {https://doi.org/10.1103/PhysRevLett.131.169001} {\bibfield  {journal}
  {\bibinfo  {journal} {Phys. Rev. Lett.}\ }\textbf {\bibinfo {volume} {131}},\
  \bibinfo {pages} {169001} (\bibinfo {year} {2023})},\ \Eprint
  {https://arxiv.org/abs/2310.13869} {arXiv:2310.13869 [astro-ph.HE]}
  \BibitemShut {NoStop}%
\bibitem [{\citenamefont {Cotesta}\ \emph {et~al.}(2022)\citenamefont
  {Cotesta}, \citenamefont {Carullo}, \citenamefont {Berti},\ and\
  \citenamefont {Cardoso}}]{Cotesta:2022pci}%
  \BibitemOpen
  \bibfield  {author} {\bibinfo {author} {\bibfnamefont {R.}~\bibnamefont
  {Cotesta}}, \bibinfo {author} {\bibfnamefont {G.}~\bibnamefont {Carullo}},
  \bibinfo {author} {\bibfnamefont {E.}~\bibnamefont {Berti}},\ and\ \bibinfo
  {author} {\bibfnamefont {V.}~\bibnamefont {Cardoso}},\ }\href
  {https://doi.org/10.1103/PhysRevLett.129.111102} {\bibfield  {journal}
  {\bibinfo  {journal} {Phys. Rev. Lett.}\ }\textbf {\bibinfo {volume} {129}},\
  \bibinfo {pages} {111102} (\bibinfo {year} {2022})},\ \Eprint
  {https://arxiv.org/abs/2201.00822} {arXiv:2201.00822 [gr-qc]} \BibitemShut
  {NoStop}%
\bibitem [{\citenamefont {Carullo}\ \emph {et~al.}(2023)\citenamefont
  {Carullo}, \citenamefont {Del~Pozzo},\ and\ \citenamefont {Veitch}}]{pyRing}%
  \BibitemOpen
  \bibfield  {author} {\bibinfo {author} {\bibfnamefont {G.}~\bibnamefont
  {Carullo}}, \bibinfo {author} {\bibfnamefont {W.}~\bibnamefont {Del~Pozzo}},\
  and\ \bibinfo {author} {\bibfnamefont {J.}~\bibnamefont {Veitch}},\ }\href
  {https://doi.org/10.5281/zenodo.8165508} {\bibinfo {title} {\texttt{pyRing}:
  a time-domain ringdown analysis python package}},\ \bibinfo {howpublished}
  {\href{https://git.ligo.org/lscsoft/pyring}{git.ligo.org/lscsoft/pyring}}
  (\bibinfo {year} {2023})\BibitemShut {NoStop}%
\bibitem [{\citenamefont {Isi}\ and\ \citenamefont {Farr}(2022)}]{Isi:2022mhy}%
  \BibitemOpen
  \bibfield  {author} {\bibinfo {author} {\bibfnamefont {M.}~\bibnamefont
  {Isi}}\ and\ \bibinfo {author} {\bibfnamefont {W.~M.}\ \bibnamefont {Farr}},\
  }\href@noop {} {\  (\bibinfo {year} {2022})},\ \Eprint
  {https://arxiv.org/abs/2202.02941} {arXiv:2202.02941 [gr-qc]} \BibitemShut
  {NoStop}%
\bibitem [{\citenamefont {Finch}\ and\ \citenamefont
  {Moore}(2022)}]{Finch:2022ynt}%
  \BibitemOpen
  \bibfield  {author} {\bibinfo {author} {\bibfnamefont {E.}~\bibnamefont
  {Finch}}\ and\ \bibinfo {author} {\bibfnamefont {C.~J.}\ \bibnamefont
  {Moore}},\ }\href {https://doi.org/10.1103/PhysRevD.106.043005} {\bibfield
  {journal} {\bibinfo  {journal} {Phys. Rev. D}\ }\textbf {\bibinfo {volume}
  {106}},\ \bibinfo {pages} {043005} (\bibinfo {year} {2022})},\ \Eprint
  {https://arxiv.org/abs/2205.07809} {arXiv:2205.07809 [gr-qc]} \BibitemShut
  {NoStop}%
\bibitem [{\citenamefont {Crisostomi}\ \emph {et~al.}(2023)\citenamefont
  {Crisostomi}, \citenamefont {Dey}, \citenamefont {Barausse},\ and\
  \citenamefont {Trotta}}]{Crisostomi:2023tle}%
  \BibitemOpen
  \bibfield  {author} {\bibinfo {author} {\bibfnamefont {M.}~\bibnamefont
  {Crisostomi}}, \bibinfo {author} {\bibfnamefont {K.}~\bibnamefont {Dey}},
  \bibinfo {author} {\bibfnamefont {E.}~\bibnamefont {Barausse}},\ and\
  \bibinfo {author} {\bibfnamefont {R.}~\bibnamefont {Trotta}},\ }\href
  {https://doi.org/10.1103/PhysRevD.108.044029} {\bibfield  {journal} {\bibinfo
   {journal} {Phys. Rev. D}\ }\textbf {\bibinfo {volume} {108}},\ \bibinfo
  {pages} {044029} (\bibinfo {year} {2023})},\ \Eprint
  {https://arxiv.org/abs/2305.18528} {arXiv:2305.18528 [gr-qc]} \BibitemShut
  {NoStop}%
\bibitem [{\citenamefont {Isi}\ \emph {et~al.}(2019)\citenamefont {Isi},
  \citenamefont {Giesler}, \citenamefont {Farr}, \citenamefont {Scheel},\ and\
  \citenamefont {Teukolsky}}]{Isi:2019aib}%
  \BibitemOpen
  \bibfield  {author} {\bibinfo {author} {\bibfnamefont {M.}~\bibnamefont
  {Isi}}, \bibinfo {author} {\bibfnamefont {M.}~\bibnamefont {Giesler}},
  \bibinfo {author} {\bibfnamefont {W.~M.}\ \bibnamefont {Farr}}, \bibinfo
  {author} {\bibfnamefont {M.~A.}\ \bibnamefont {Scheel}},\ and\ \bibinfo
  {author} {\bibfnamefont {S.~A.}\ \bibnamefont {Teukolsky}},\ }\href
  {https://doi.org/10.1103/PhysRevLett.123.111102} {\bibfield  {journal}
  {\bibinfo  {journal} {Phys. Rev. Lett.}\ }\textbf {\bibinfo {volume} {123}},\
  \bibinfo {pages} {111102} (\bibinfo {year} {2019})},\ \Eprint
  {https://arxiv.org/abs/1905.00869} {arXiv:1905.00869 [gr-qc]} \BibitemShut
  {NoStop}%
\bibitem [{\citenamefont {Carullo}\ \emph {et~al.}(2019)\citenamefont
  {Carullo}, \citenamefont {Del~Pozzo},\ and\ \citenamefont
  {Veitch}}]{Carullo:2019flw}%
  \BibitemOpen
  \bibfield  {author} {\bibinfo {author} {\bibfnamefont {G.}~\bibnamefont
  {Carullo}}, \bibinfo {author} {\bibfnamefont {W.}~\bibnamefont {Del~Pozzo}},\
  and\ \bibinfo {author} {\bibfnamefont {J.}~\bibnamefont {Veitch}},\ }\href
  {https://doi.org/10.1103/PhysRevD.99.123029} {\bibfield  {journal} {\bibinfo
  {journal} {Phys. Rev. D}\ }\textbf {\bibinfo {volume} {99}},\ \bibinfo
  {pages} {123029} (\bibinfo {year} {2019})},\ \bibinfo {note} {[Erratum:
  Phys.Rev.D 100, 089903 (2019)]},\ \Eprint {https://arxiv.org/abs/1902.07527}
  {arXiv:1902.07527 [gr-qc]} \BibitemShut {NoStop}%
\bibitem [{\citenamefont {Siegel}\ \emph {et~al.}(2023)\citenamefont {Siegel},
  \citenamefont {Isi},\ and\ \citenamefont {Farr}}]{Siegel:2023lxl}%
  \BibitemOpen
  \bibfield  {author} {\bibinfo {author} {\bibfnamefont {H.}~\bibnamefont
  {Siegel}}, \bibinfo {author} {\bibfnamefont {M.}~\bibnamefont {Isi}},\ and\
  \bibinfo {author} {\bibfnamefont {W.~M.}\ \bibnamefont {Farr}},\ }\href
  {https://doi.org/10.1103/PhysRevD.108.064008} {\bibfield  {journal} {\bibinfo
   {journal} {Phys. Rev. D}\ }\textbf {\bibinfo {volume} {108}},\ \bibinfo
  {pages} {064008} (\bibinfo {year} {2023})},\ \Eprint
  {https://arxiv.org/abs/2307.11975} {arXiv:2307.11975 [gr-qc]} \BibitemShut
  {NoStop}%
\bibitem [{\citenamefont {Wang}\ \emph {et~al.}(2023)\citenamefont {Wang},
  \citenamefont {Capano}, \citenamefont {Abedi}, \citenamefont {Kastha},
  \citenamefont {Krishnan}, \citenamefont {Nielsen}, \citenamefont {Nitz},\
  and\ \citenamefont {Westerweck}}]{wang2023frequencydomain}%
  \BibitemOpen
  \bibfield  {author} {\bibinfo {author} {\bibfnamefont {Y.-F.}\ \bibnamefont
  {Wang}}, \bibinfo {author} {\bibfnamefont {C.~D.}\ \bibnamefont {Capano}},
  \bibinfo {author} {\bibfnamefont {J.}~\bibnamefont {Abedi}}, \bibinfo
  {author} {\bibfnamefont {S.}~\bibnamefont {Kastha}}, \bibinfo {author}
  {\bibfnamefont {B.}~\bibnamefont {Krishnan}}, \bibinfo {author}
  {\bibfnamefont {A.~B.}\ \bibnamefont {Nielsen}}, \bibinfo {author}
  {\bibfnamefont {A.~H.}\ \bibnamefont {Nitz}},\ and\ \bibinfo {author}
  {\bibfnamefont {J.}~\bibnamefont {Westerweck}},\ }\href@noop {} {\bibinfo
  {title} {A frequency-domain perspective on gw150914 ringdown overtone}}
  (\bibinfo {year} {2023}),\ \Eprint {https://arxiv.org/abs/2310.19645}
  {arXiv:2310.19645 [gr-qc]} \BibitemShut {NoStop}%
\bibitem [{\citenamefont {Farr}()}]{pyRing_issue}%
  \BibitemOpen
  \bibfield  {author} {\bibinfo {author} {\bibfnamefont {W.}~\bibnamefont
  {Farr}},\ }\href@noop {} {}\bibinfo {howpublished}
  {\url{https://git.ligo.org/lscsoft/pyring/-/issues/7}}\BibitemShut {NoStop}%
\bibitem [{\citenamefont {Abbott}\ \emph {et~al.}(2021)\citenamefont {Abbott}
  \emph {et~al.}}]{LIGOScientific:2020tif}%
  \BibitemOpen
  \bibfield  {author} {\bibinfo {author} {\bibfnamefont {R.}~\bibnamefont
  {Abbott}} \emph {et~al.} (\bibinfo {collaboration} {LIGO Scientific,
  Virgo}),\ }\href {https://doi.org/10.1103/PhysRevD.103.122002} {\bibfield
  {journal} {\bibinfo  {journal} {Phys. Rev. D}\ }\textbf {\bibinfo {volume}
  {103}},\ \bibinfo {pages} {122002} (\bibinfo {year} {2021})},\ \Eprint
  {https://arxiv.org/abs/2010.14529} {arXiv:2010.14529 [gr-qc]} \BibitemShut
  {NoStop}%
\bibitem [{\citenamefont {Baibhav}\ \emph {et~al.}(2023)\citenamefont
  {Baibhav}, \citenamefont {Cheung}, \citenamefont {Berti}, \citenamefont
  {Cardoso}, \citenamefont {Carullo}, \citenamefont {Cotesta}, \citenamefont
  {Del~Pozzo},\ and\ \citenamefont {Duque}}]{Baibhav:2023clw}%
  \BibitemOpen
  \bibfield  {author} {\bibinfo {author} {\bibfnamefont {V.}~\bibnamefont
  {Baibhav}}, \bibinfo {author} {\bibfnamefont {M.~H.-Y.}\ \bibnamefont
  {Cheung}}, \bibinfo {author} {\bibfnamefont {E.}~\bibnamefont {Berti}},
  \bibinfo {author} {\bibfnamefont {V.}~\bibnamefont {Cardoso}}, \bibinfo
  {author} {\bibfnamefont {G.}~\bibnamefont {Carullo}}, \bibinfo {author}
  {\bibfnamefont {R.}~\bibnamefont {Cotesta}}, \bibinfo {author} {\bibfnamefont
  {W.}~\bibnamefont {Del~Pozzo}},\ and\ \bibinfo {author} {\bibfnamefont
  {F.}~\bibnamefont {Duque}},\ }\href@noop {} {\  (\bibinfo {year} {2023})},\
  \Eprint {https://arxiv.org/abs/2302.03050} {arXiv:2302.03050 [gr-qc]}
  \BibitemShut {NoStop}%
\bibitem [{\citenamefont {Nee}\ \emph {et~al.}(2023)\citenamefont {Nee},
  \citenamefont {V\"olkel},\ and\ \citenamefont {Pfeiffer}}]{Nee:2023osy}%
  \BibitemOpen
  \bibfield  {author} {\bibinfo {author} {\bibfnamefont {P.~J.}\ \bibnamefont
  {Nee}}, \bibinfo {author} {\bibfnamefont {S.~H.}\ \bibnamefont {V\"olkel}},\
  and\ \bibinfo {author} {\bibfnamefont {H.~P.}\ \bibnamefont {Pfeiffer}},\
  }\href@noop {} {\  (\bibinfo {year} {2023})},\ \Eprint
  {https://arxiv.org/abs/2302.06634} {arXiv:2302.06634 [gr-qc]} \BibitemShut
  {NoStop}%
\bibitem [{\citenamefont {{Ho-Yeuk Cheung}}\ \emph {et~al.}(2023)\citenamefont
  {{Ho-Yeuk Cheung}}, \citenamefont {{Berti}}, \citenamefont {{Baibhav}},\ and\
  \citenamefont {{Cotesta}}}]{2023arXiv231004489H}%
  \BibitemOpen
  \bibfield  {author} {\bibinfo {author} {\bibfnamefont {M.}~\bibnamefont
  {{Ho-Yeuk Cheung}}}, \bibinfo {author} {\bibfnamefont {E.}~\bibnamefont
  {{Berti}}}, \bibinfo {author} {\bibfnamefont {V.}~\bibnamefont {{Baibhav}}},\
  and\ \bibinfo {author} {\bibfnamefont {R.}~\bibnamefont {{Cotesta}}},\ }\href
  {https://doi.org/10.48550/arXiv.2310.04489} {\bibfield  {journal} {\bibinfo
  {journal} {arXiv e-prints}\ ,\ \bibinfo {eid} {arXiv:2310.04489}} (\bibinfo
  {year} {2023})},\ \Eprint {https://arxiv.org/abs/2310.04489}
  {arXiv:2310.04489 [gr-qc]} \BibitemShut {NoStop}%
\bibitem [{\citenamefont {Cheung}\ \emph {et~al.}(2023)\citenamefont {Cheung}
  \emph {et~al.}}]{Cheung:2022rbm}%
  \BibitemOpen
  \bibfield  {author} {\bibinfo {author} {\bibfnamefont {M.~H.-Y.}\
  \bibnamefont {Cheung}} \emph {et~al.},\ }\href
  {https://doi.org/10.1103/PhysRevLett.130.081401} {\bibfield  {journal}
  {\bibinfo  {journal} {Phys. Rev. Lett.}\ }\textbf {\bibinfo {volume} {130}},\
  \bibinfo {pages} {081401} (\bibinfo {year} {2023})},\ \Eprint
  {https://arxiv.org/abs/2208.07374} {arXiv:2208.07374 [gr-qc]} \BibitemShut
  {NoStop}%
\bibitem [{\citenamefont {Mitman}\ \emph {et~al.}(2023)\citenamefont {Mitman}
  \emph {et~al.}}]{Mitman:2022qdl}%
  \BibitemOpen
  \bibfield  {author} {\bibinfo {author} {\bibfnamefont {K.}~\bibnamefont
  {Mitman}} \emph {et~al.},\ }\href
  {https://doi.org/10.1103/PhysRevLett.130.081402} {\bibfield  {journal}
  {\bibinfo  {journal} {Phys. Rev. Lett.}\ }\textbf {\bibinfo {volume} {130}},\
  \bibinfo {pages} {081402} (\bibinfo {year} {2023})},\ \Eprint
  {https://arxiv.org/abs/2208.07380} {arXiv:2208.07380 [gr-qc]} \BibitemShut
  {NoStop}%
\bibitem [{\citenamefont {Khera}\ \emph {et~al.}(2023)\citenamefont {Khera},
  \citenamefont {Ribes~Metidieri}, \citenamefont {Bonga}, \citenamefont
  {Forteza}, \citenamefont {Krishnan}, \citenamefont {Poisson}, \citenamefont
  {Pook-Kolb}, \citenamefont {Schnetter},\ and\ \citenamefont
  {Yang}}]{Khera:2023lnc}%
  \BibitemOpen
  \bibfield  {author} {\bibinfo {author} {\bibfnamefont {N.}~\bibnamefont
  {Khera}}, \bibinfo {author} {\bibfnamefont {A.}~\bibnamefont
  {Ribes~Metidieri}}, \bibinfo {author} {\bibfnamefont {B.}~\bibnamefont
  {Bonga}}, \bibinfo {author} {\bibfnamefont {X.~J.}\ \bibnamefont {Forteza}},
  \bibinfo {author} {\bibfnamefont {B.}~\bibnamefont {Krishnan}}, \bibinfo
  {author} {\bibfnamefont {E.}~\bibnamefont {Poisson}}, \bibinfo {author}
  {\bibfnamefont {D.}~\bibnamefont {Pook-Kolb}}, \bibinfo {author}
  {\bibfnamefont {E.}~\bibnamefont {Schnetter}},\ and\ \bibinfo {author}
  {\bibfnamefont {H.}~\bibnamefont {Yang}},\ }\href@noop {} {\  (\bibinfo
  {year} {2023})},\ \Eprint {https://arxiv.org/abs/2306.11142}
  {arXiv:2306.11142 [gr-qc]} \BibitemShut {NoStop}%
\bibitem [{\citenamefont {Abbott}\ \emph {et~al.}(2016)\citenamefont {Abbott}
  \emph {et~al.}}]{LIGOScientific:2016lio}%
  \BibitemOpen
  \bibfield  {author} {\bibinfo {author} {\bibfnamefont {B.~P.}\ \bibnamefont
  {Abbott}} \emph {et~al.} (\bibinfo {collaboration} {LIGO Scientific,
  Virgo}),\ }\href {https://doi.org/10.1103/PhysRevLett.116.221101} {\bibfield
  {journal} {\bibinfo  {journal} {Phys. Rev. Lett.}\ }\textbf {\bibinfo
  {volume} {116}},\ \bibinfo {pages} {221101} (\bibinfo {year} {2016})},\
  \bibinfo {note} {[Erratum: Phys.Rev.Lett. 121, 129902 (2018)]},\ \Eprint
  {https://arxiv.org/abs/1602.03841} {arXiv:1602.03841 [gr-qc]} \BibitemShut
  {NoStop}%
\end{thebibliography}%

\end{document}